\documentclass[twocolumn,prb]{revtex4}
\usepackage{graphicx}
\usepackage{bm}
\usepackage{amsmath}

\begin{document}

\title{Theory of Spin Transport Induced by Ferromagnetic Proximity On
a Two-Dimensional Electron Gas}

\author{J. P. McGuire, C. Ciuti, L. J. Sham}

\affiliation{Department of Physics, University of California San
  Diego, La Jolla, CA 92093-0319.}

\begin{abstract}
A theory of the proximity effects of the exchange splitting in a
ferromagnetic metal on a two dimensional electron gas (2DEG) in a
semiconductor is presented.  The resulting spin-dependent energy and
lifetime in the 2DEG create a marked spin-splitting in the driven
in-plane current. The theory of the planar transport allows for
current leakage into the ferromagnetic layer through the interface,
which leads to a competition between drift and diffusion.  The
spin-dependent in-plane conductivity of the 2DEG may be exploited to
provide a new paradigm for spintronics devices based on planar devices
in a field-effect transistor configuration.  An illustrative example
is provided through the transport theory of a proposed spin-valve
which consists of a field-effect transistor configuration with two
ferromagnetic gates.  Results are provided for two experimentally
accessible systems: the silicon inversion layer and the 
naturally-formed InAs accumulation layer.
\end{abstract}


\date{\today}

\maketitle

\section{Introduction}
The emerging field of spintronics aims to implement semiconductor
devices which utilize both the carrier charge and spin degrees of
freedom. \cite{spinreview} Research in the field has been inspired by
the early device proposal of Datta and Das, \cite{datta} which
consists of spin injection through a ferromagnetic-semiconductor
interface and spin manipulation using the Rashba spin-orbit
effect. \cite{rashba} There has been recent progress in achieving
maximum spin injection \cite{fiederling, ohno, zhu, hanbicki} and in
characterization of the Rashba effect \cite{nitta1, nitta2, marcus1,
marcus2}. However, a spin device based on injection has a high tunnel
resistance, analogous to a Schottky diode. We have suggested an
alternate approach of spin creation and manipulation which makes use
of the proximity effects of a ferromagnet on a
semiconductor. \cite{ciuti2} The eventual device would avoid the high
tunnelling resistance by keeping the main driven current path entirely
in the semiconductor through normal ohmic nonmagnetic leads,
resembling the field-effect transistor design.

In this paper we present a comprehensive theory of the consequences of
the ferromagnetic proximity on the equilibrium and transport
properties of the electrons in the semiconductor.  First, we fully
examine the coupling of a semiconductor two dimensional electron gas
(2DEG) with a ferromagnetic layer through a very thin potential
barrier. Explicit calculations are provided for two realizable
systems, the silicon inversion layer and the naturally-formed InAs
surface layer, both with ferromagnetic gates.  The coupling is
conveniently treated by a Green's function method which can account
for realistic confinement for the semiconductor 2DEG as well as the
effects of a short electron mean free path in the ferromagnet.  The
complex self-energy of the 2DEG due to the interaction with the
ferromagnet contains both static and dynamic effects (a Zeeman-like
splitting and spin-dependent scattering times, respectively).  Both
properties alter the in-plane conductivity and can be exploited
through the spin-dependent transport under the ferromagnetic gate.  As
an illustration of the possible consequences of the ferromagnetic
proximity on the semiconductor, we show in detail the behavior of the
density and current in a previously proposed \cite{ciuti2} planar
spin-valve in MOSFET-style configuration with two adjacent
ferromagnetic gates with reversible magnetizations.  We find that the
proposed device has a significant magnetoresistance effect for
reasonable system parameters.

The coupling of the 2DEG to the ferromagnetic gate exponentially
decreases with their separation by a potential barrier.  Our study
includes oxide barriers down to the smallest state-of-the-art
thickness and direct contact between the 2DEG and the ferromagnet to
ensure sizable spin effects.  A consequence of this requirement for
ultrathin oxides is that current leakage into the ferromagnetic
gate(s) becomes an integral part of the in-plane electron transport.
This naturally results in the need to consider both the current driven
by the in-plane electric field and the diffusion driven by the density
gradient, analogous to the bulk case. \cite{flatte1,flatte2} Although
such leakage effects would decrease the efficiency of normal
field-effect transistors, we find that they actually enhance the
spin-dependence of the electron transport from the source to the
drain.

We argue that the relevance of proximity effects between a
semiconductor and a ferromagnet is supported by recent experiments.  A
series of optical Faraday rotation experiments \cite{kawakami,
epstein} have shown that unpolarized non-equilibrium electrons in a
semiconductor can spontaneously acquire a net spin polarization in the
presence of a ferromagnetic interface. We have interpreted
\cite{ciuti1} the observed effect as arising from the spin-dependent
reflection of electrons off the ferromagnetic interface.  Assuming
negligible spin scattering at the interface, the spontaneous spin
polarization produced in the semiconductor is indicative of the
strength of the coupling across the semiconductor-ferromagnet
junction.  In addition, replacing the Schottky barrier \cite{kawakami,
epstein} with a much thinner oxide will only increase the interaction.

Proximity effects between dissimilar materials have been used to
describe how the ordered state of one medium induces a similar order
in the other medium which is otherwise normal.  The induced order
parameter decays away from the interface.  The common examples are the
proximity effects between a superconductor and a normal metal (or a
semiconductor 2DEG) and between a ferromagnet and a paramagnetic
metal.  The induced ferromagnetic order in a non-magnetic metal is
weak and requires the metal to be superparamagnetic, \cite{ivan} such
as Pd or Pt.  In this paper we concentrate on the proximity effect
between a ferromagnet and a semiconductor 2DEG. The 2DEG is examined
rather than the bulk semiconductor because it has two advantages: (1)
the 2DEG is confined near the interface where it is more susceptible
to the influence of the exchange-split electrons in the ferromagnet,
and (2) while the induced polarization in the 2DEG may be too small
for a magnetization measurement, the influence on the spin-polarized
transport is the ultimate goal.  Moreover, there exists a large amount
of knowledge and technology dedicated to
the manipulation of
semiconductor 2DEG's, further increasing the potential for spintronics
devices.

In Sec. \ref{model_of_structure} we explain the model of the
semiconductor 2DEG-ferromagnet used in the calculations, followed by
the derivation of an effective tight-binding Hamiltonian
(Sec. \ref{derivation_of_eff_H}) which is used in a Green's function
formalism (Sec. \ref{greens_functions}) to approximate the coupling
between the semiconductor 2DEG and the ferromagnet.  In
Sec. \ref{coupling_results} we calculate the self-energy of the 2DEG
coupled to the ferromagnet for two experimentally realizable systems,
the silicon inversion layer (Sec. \ref{silicon_coupling}) and the
naturally-formed InAs surface layer (Sec. \ref{inas_coupling}).  In
Sec. \ref{spin_dependent_transport} we derive the equations and show
representative results for the 2DEG density and in-plane current for
the silicon inversion layer (Sec. \ref{si_transport}) and for the InAs
surface layer (Sec. \ref{inas_transport}).  In
Sec. \ref{spin_valve_two_gates} we calculate the 2DEG density and
current for a spin-valve with two ferromagnetic gates for the silicon
(Sec. \ref{si_spin_valve}) and the InAs (Sec. \ref{inas_spin_valve})
systems.  We summarize our findings and draw conclusions in
Sec. \ref{conc}.

\section{Model of the Electronic Structure}
\label{model_of_structure}
\begin{figure}[t!]
\includegraphics[width=8cm]{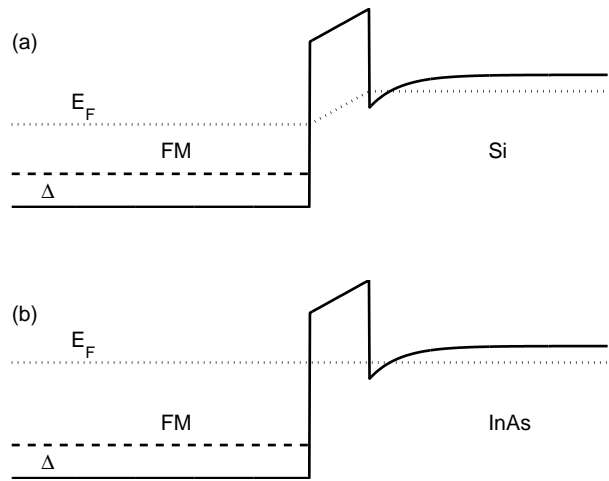}
\caption{Band diagrams for (a) the silicon inversion layer and (b) the
  InAs surface layer, both separated from a ferromagnet by a thin
  oxide barrier.  The ferromagnet, with exchange splitting $\Delta$,
  is on the left and the semiconductor 2DEG is on the right.  In (a),
  the Fermi level in the ferromagnet is lower than the Fermi level in
  the silicon inversion layer. In (b), the Fermi level in the
  ferromagnet is equal to the Fermi level in the InAs surface layer.}
  \label{band_diagrams}
\end{figure}

The two particular systems we will consider have band diagrams as
shown in Fig. \ref{band_diagrams}.  The semiconductor 2DEG is on the
right, which is induced by a gate bias for the silicon inversion layer
(Fig.~\ref{band_diagrams}(a)) and is naturally formed for the InAs
surface layer (Fig.~\ref{band_diagrams}(b)).  The left side is the
ferromagnet, which is modelled as parabolic bands split by the
exchange energy $\Delta$.\cite{slon} A very thin oxide separates the
2DEG from the ferromagnet.  This oxide is necessary in the silicon
system because a gate bias is necessary for the creation of the
inversion layer.  Because the surface layer in the InAs system forms
naturally, no gate bias is necessary and hence intimate contact (no
oxide barrier) is possible between the InAs surface layer and the
ferromagnet.

All calculations are done within the effective mass approximation.
This, along with the exchange-split parabolic band approximation,
cannot possibly account for all band effects in the ferromagnet.  We
wish to show in a transparent way how the spin-dependence in the
ferromagnet can influence the 2DEG; more realistic calculations will
be necessary for comparison with experiment, but the essential effects
we deduce here should remain valid. In addition, the two spin channels
are considered to be completely decoupled and are labelled by $ + $
and $ - $.  Experimentally, it appears that the spin lifetime in
semiconductor heterostructures can persist for long times and over
long distances, including through heterostructure
interfaces.\cite{kikkawa1, kikkawa2, kikkawa3, malajovich} We neglect
any mixing due to the spin-orbit effect or other spin-flip processes.
The neglect of the Rashba spin-orbit interaction \cite{rashba} is
valid in Si and holds less well in the InAs surface layer, but
sufficiently well without gate voltage. \cite{nitta2}

In a recent article\cite{ciuti2} we reported the calculation of the
coupling between a silicon inversion layer and a ferromagnetic gate in
the effective mass approximation using a triangular potential in the
semiconductor region.  The simplicity of the potential allowed us to
use the wavefunction matching conditions to derive an equation
specifying the complex energy of the eigenstate of the coupled system.
Strong scattering in the ferromagnet was incorporated by putting a
phenomenological damping into the wavefunction in the ferromagnet
region.  The triangular potential approximation, although valid in the
silicon system for strong inversion,\cite{ando} is not appropriate for
the weakly confined 2DEG at the InAs surface. For this reason, the
InAs potential is calculated self-consisently with the surface layer
density distribution using the coupled Schrodinger and Poisson
equations.  A Green's function method is devoloped to calculate the
coupling between the semiconductor 2DEG and the ferromagnet which
works for both the triangle potential in the silicon system and the
self-consistent potential in the InAs system.  In addition, the
Green's function approach is ideal because it allows us to include
scattering in a more natural and rigorous manner than in the wave
function approach.  In brief, the coupled semiconductor
2DEG-ferromagnet system is solved in three steps. (1) Separate the
semiconductor and ferromagnet regions and solve them separately.  (2)
Approximate the coupling between the two subsystems by transforming
the original effective mass Hamiltonian into a ``tight-binding''
(i.e., tunneling) Hamiltonian.  (3) Calculate the self-energy of 2DEG
electrons due to the interaction with the ferromagnet.

\subsection{Derivation of the Effective Hamiltonian}
\label{derivation_of_eff_H}
The effective mass Hamiltonian for the ferromagnet-oxide-semiconductor
junction (see Fig. \ref{band_diagrams}),
\begin{eqnarray}
\nonumber H &=& \frac{-\hbar^2}{2} \frac{d}{dz} \left (
\frac{1}{m^\star(z)} \frac{d}{dz} \right )\\ \nonumber &~& + \left (
U_{\text{fm}} + \frac{\Delta}{2} {\bm \sigma} \cdot \hat{\bf M} \right
) \Theta(-z) \\&~& + U_{\text{b}}(z) \Theta (z_{\rm b}-z) \Theta (z) +
U_{\rm sc}(z) \Theta (z-z_{\rm b})~,
\label{eff_mass_H}
\end{eqnarray}
can be written as $H=K+V^{\text{sc}}+V^{\text{fm}}_\sigma$, where $K$
is the kinetic energy and the {\it semiconductor} and {\it
ferromagnet} potentials are defined respectively as
\begin{eqnarray}
\nonumber V^{\text{sc}}(z) &=& U_{\text{sc}}(z) \Theta
(z-z_{\text{b}}) + 0 \cdot \Theta (z_{\text{b}}-z)~, \\ \nonumber
V^{\text{fm}}_{\sigma} (z) &=& U_{\text{b}}(z) \Theta (z_{\text{b}}-z)
\Theta (z) + 0 \cdot \Theta(z-z_{\text{b}}) \\ &~~~~& + \left (
U_{\text{fm}} + \frac{\Delta}{2} {\bm \sigma} \cdot \hat{\bf M} \right
) \Theta(-z)~.
\label{split_pot}
\end{eqnarray}
The zero of energy is put at the right side of the barrier for
convenience.  Basically, $ V^{\text{fm}}_\sigma(z) $ consists of the
spin-dependent ferromagnet potential $ \left ( U_{\text{fm}} +
\frac{\Delta}{2} {\bm \sigma} \cdot\hat{ {\bf M}} \right ) $ for
$z<0$, the oxide barrier potential $ U_{\text{b}}(z) $ for
$0<z<z_{\text{b}}$, and zero for $z>z_{\text{b}}$.  From now on we
choose the spin quantization axis to be parallel to the magnetization
$\bf{M}$ in the ferromagnet, so that $ V^{\text{fm}}_\sigma
\rightarrow V^{\text{fm}}_\pm $.  The semiconductor potential $
V^{\text{sc}}(z) $ consists of the 2DEG confinement potential $
U_{\text{sc}}(z) $ for $ z>z_{\text{b}}$ and is zero everywhere else.

\begin{figure}
\includegraphics[width=8cm]{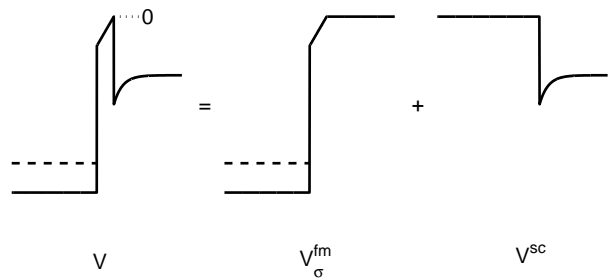}
\caption{A cartoon of the splitting up of the potential.  The original
  band diagram is on the left, and this is split into a {\it
  ferromagnetic} potential $V^{\text{fm}}_\sigma$ and a {\it
  semiconductor} potential $V^{\text{sc}}$.} \label{coupling}
\end{figure}

This separation of the potential (illustrated in Fig. \ref{coupling})
is convenient because it allows us to solve the two subsystems with
potentials $ V^{\text{fm}}_{\pm}(z) $ and $ V^{\text{sc}}(z) $
separately, and then calculate the coupling between them.  The
subsystems are solved, yielding the semiconductor eigenstates $ \left
\{ \chi^{\text{sc}}_n (z) \right \}$ with energy
$\epsilon^{\text{sc}}_n$ and the ferromagnet eigenstates $ \left \{
\chi^{\text{fm}}_{k,\pm} (z) \right \} $ with energy
$\epsilon^{\text{fm}}_{k,\pm}$.  The finite overlap between the
wavefunctions for the two subsystems implies that they are not
mutually orthogonal.  To ensure proper Fermionic anti-commutation
relations between the operators in all parts of the system, we
orthogonalize the ferromagnetic eigenstates to the discrete 2DEG
eigenstates,
\begin{equation}
\phi^{\text{fm}}_{k,\pm}(z) = \chi^{\text{fm}}_{k,\pm}(z)-\sum_n
\chi^{\text{sc}}_n (z) \langle \chi^{\text{sc}}_n |
\chi^{\text{fm}}_{k,\pm} \rangle ~.
\label{orthogonal}
\end{equation}
This allows us to write a general state of the system as
\begin{equation}
\psi_\pm (z) = \sum_n \chi^{\text{sc}}_n (z) a_{n,\pm} + \sum_k
\phi^{\text{fm}}_{k,\pm} (z) c_{k,\pm}~,
\label{wavefunction}
\end{equation}
where $a_{n,\pm}$ and $c_{k,\pm}$ are the Fermion operators for the
2DEG state $\chi^{\text{sc}}_n (z) $ and the orthogonalized
ferromagnet state $\phi^{\text{fm}}_{k,\pm}(z)$ respectively.  The
only Fermion operators that do not anticommute properly are
\begin{eqnarray}
\nonumber \left \{ c_{k,\pm} , c_{k^\prime,\pm}^\dagger \right \}& =&
\delta_{k,k^\prime} - 2 \langle \chi^{\text{sc}}_n |
\chi^{\text{fm}}_{k,\pm} \rangle \langle
\chi^{\text{fm}}_{k^\prime,\pm} | \chi^{\text{sc}}_n \rangle \\
&\approx& \delta_{k,k^\prime}~.
\label{anticommutator}
\end{eqnarray}
As an approximation we keep only terms that have a single
exponentially decaying term in an overlap integral.  Since $ \langle
\chi^{\text{sc}}_n | \chi^{\text{fm}}_{k,\pm} \rangle \langle
\chi^{\text{fm}}_{k^\prime,\pm} | \chi^{\text{sc}}_n \rangle $
contains the product of two overlap integrals with an exponentially
decaying term, we neglect it, leaving an approximate set of Fermionic
operators.

Now we must calculate all the matrix elements involved in $\langle
\psi_\pm | K+V^{\text{fm}}_\pm+V^{\text{sc}} | \psi_\pm \rangle$,
keeping the approximation as mentioned above.  The matrix elements
between different semiconductor states are
\begin{eqnarray}
\nonumber \langle \chi^{\text{sc}}_{n^\prime} | K+V^{\text{fm}}_\pm +
V^{\text{sc}} | \chi^{\text{sc}}_n \rangle &=& \epsilon^{\text{sc}}_n
\delta_{n,n^\prime} + V^{\text{fm}}_{n^\prime,n,\pm} \\ &\approx&
\epsilon^{\text{sc}}_n \delta_{n,n^\prime}~,
\label{chi_chi}
\end{eqnarray}
where
\begin{equation}
V^{\text{fm}}_{n^\prime,n,\pm} = \int dz \left (
\chi^{\text{sc}}_{n^\prime} \right )^\star V^{\text{fm}}_\pm
\chi^{\text{sc}}_n
\label{V_n_n}
\end{equation}
is neglected because the integral contains two exponentially decaying
functions.  The matrix elements between a semiconductor state and a
ferromagnet state are
\begin{eqnarray}
\nonumber \langle \phi^{\text{fm}}_{k,\pm} | K + V^{\text{fm}}_\pm +
V^{\text{sc}} | \chi^{\text{sc}}_{n,\pm} \rangle &=&
\epsilon^{\text{sc}}_n \langle \phi^{\text{fm}}_{k,\pm} |
\chi^{\text{sc}}_n \rangle + V^{\text{fm}}_{k,n,\pm} \\ &=&
V^{\text{fm}}_{k,n,\pm}~,
\label{phi_chi}
\end{eqnarray}
where $ \langle \phi^{\text{fm}}_{k,\pm} | \chi^{\text{sc}}_n \rangle
=0 $ because of the orthogonality between the semiconductor and
ferromagnet states, and
\begin{equation}
V^{\text{fm}}_{k,n,\pm} = \int dz \left ( \phi^{\text{fm}}_{k,\pm} \right
)^\star V^{\text{fm}}_\pm \chi^{\text{sc}}_n
\label{V_k_n}
\end{equation}
is kept because there is only a single exponentially decaying
wavefunction in the nonzero integration region.  Finally, the matrix
elements between ferromagnetic states are
\begin{eqnarray}
\nonumber \langle \phi^{\text{fm}}_{k,\pm} | K + V^{\text{fm}}_\pm +
V^{\text{sc}} | \phi^{\text{fm}}_{k^\prime,\pm} \rangle &=&
\epsilon^{\text{fm}}_{k,\pm} \delta_{k,k^\prime} + W_{k,k^\prime,\pm}
\\ &\approx& \epsilon^{\text{fm}}_{k,\pm} \delta_{k,k^\prime}
\label{phi_phi}
\end{eqnarray}
where $ \epsilon^{\text{fm}}_{k,\pm} $is the energy of the state $
\chi^{\text{fm}}_{k,\pm} $ before the orthogonalization to the 2DEG
states and
\begin{eqnarray}
\nonumber W_{k,k^\prime,\pm} &=& \langle
\chi^{\text{fm}}_{k^\prime,\pm} | V^{\text{sc}} |
\chi^{\text{fm}}_{k,\pm} \rangle \\ \nonumber &~& + \sum_n \left [
\left ( \epsilon^{\text{sc}}_n \langle \chi^{\text{fm}}_{k^\prime,\pm}
| \chi^{\text{sc}}_n \rangle + V^{\text{fm}}_{k^\prime,n,\pm} \right )
\langle \chi^{\text{sc}}_n | \chi^{\text{fm}}_{k,\pm} \rangle \right
. \\ &~& + \left .  \left ( \epsilon^{\text{sc}}_n \langle
\chi^{\text{sc}}_n | \chi^{\text{fm}}_{k,\pm} \rangle +
V^{\text{fm}}_{n,k,\pm} \right ) \langle
\chi^{\text{fm}}_{k^\prime,\pm} | \chi^{\text{sc}}_n \rangle \right ]
\label{W_k_k}
\end{eqnarray}
is neglected because all terms contain more than a single exponentially
decaying function.

We are left with an effective Hamiltonian with a simple tight-binding
form:
\begin{eqnarray}
\nonumber H &=& \sum_{n,\sigma} a^\dagger_{n,\sigma}
\epsilon^{\text{sc}}_n a_{n ,\sigma} + \sum_{k,\sigma}
c^\dagger_{k,\sigma} \epsilon^{\text{fm}}_{k ,\sigma} c_{k,\sigma} \\
&~& ~~~+ \sum_{n,k,\sigma} \left ( a^\dagger_{n ,\sigma}
V^{\text{fm}}_{n,k,\sigma} c_{k,\sigma} + c^\dagger_{k,\sigma}
V^{\text{fm}}_{k,n,\sigma} a_{n,\sigma} \right )~.
\label{tight_binding_H}
\end{eqnarray}
This Hamiltonian represents the coupling between the quantum confined
electrons in the semiconductor and the broad spin-dependent continuum 
in the ferromagnet.  We solve the Hamiltonian below using a Green's 
function approach.

\subsection{Green's functions for the Coupled System}
\label{greens_functions}
We are interested in the properties of the 2DEG electrons due to the
coupling with the ferromagnet.  The retarded Green's functions relevant to our
calculations are:
\begin{eqnarray}
G_{n,\pm}(t) &=& -i \Theta(t) \langle \left \{ a_{n,\pm}(t),
a^\dagger_{n,\pm}(0) \right \} \rangle~, \\ G_{k,\pm}(t) &=& -i
\Theta(t) \langle \left \{ c_{k ,\pm}(t), c^\dagger_{k,\pm}(0) \right
\} \rangle~, \\ G_{k,n,\pm}(t) &=& -i \Theta(t) \langle \left \{
c_{k,\pm}(t), a_{n,\pm}^\dagger (0) \right \} \rangle~,
\label{G_def}
\end{eqnarray}
where $ \Theta(t) $ ensures us that $ t > 0 $.  The unperturbed
Green's functions when the coupling $ V^{\text{fm}}_{n,k,\pm} $ is
zero are
\begin{eqnarray}
G^{(0)}_{n,\pm}(E) &=& \left ( E - \epsilon^{\text{sc}}_n + i 0^+
\right )^{-1}~, \\ G^{(0)}_{k,\pm}(E) &=& \left ( E -
\epsilon^{\text{fm}}_{k,\pm} + i \gamma^{\text{fm}}_{k,\pm} \right
)^{-1}~.
\label{G_zero}
\end{eqnarray}
Note that the ferromagnet is assumed to be a dirty conductor, so that
an imaginary part has been added to the ferromagnet energy to account
for strong spin-dependent scattering in the ferromagnet:\cite{hong} $
\epsilon^{\text{fm}}_{k,\pm} \rightarrow
\epsilon^{\text{fm}}_{k,\pm}-i \gamma^{\text{fm}}_{k,\pm} $~.

Calculating the Green's functions with the interaction from the
Hamiltonian and Fourier transforming, we have the following equation
for the full 2DEG Green's function:
\begin{equation}
G_{n, \pm}(E) = \left ( E - \epsilon^{\text{sc}}_n + i 0^+ -
\Sigma_{n,\pm}(E) \right )^{-1}~,
\label{G_n}
\end{equation}
where the self-energy of the 2DEG electrons is
\begin{equation}
\Sigma_{n,\pm}(E) = \sum_k V^{\text{fm}}_{n,k,\pm} G^{(0)}_{k,\pm}(E)
V^{\text{fm}}_{k,n,\pm}~.
\label{self_energy}
\end{equation}
The effect of the coupling on the 2DEG eigenstates is found by from
the complex energy $ \tilde{E} $ which satisfies
\begin{equation}
\tilde{E} - \epsilon^{\text{sc}}_n + i 0^+ - \Sigma_{n,\pm}(\tilde{E})
= 0~.
\label{E_tilde}
\end{equation}

The real part of the 2DEG self-energy, $ \Delta_{n,\pm}(\tilde{E}) =
{\rm Re} \left ( \Sigma_{n,\pm}(\tilde{E}) \right ) $, is the
spin-dependent level shift of the 2DEG subbands due to the coupling
with the ferromagnet.  This gives the 2DEG a static spin-splitting,
which is $ | \Delta_+ - \Delta_- | $ if only one 2DEG subband is
occupied. The two spin channels will have different densities, $ N_+
\ne N_- $.  The imaginary part of the 2DEG self-energy, $ \hbar /
2 \tau_{n,\pm} (\tilde{E}) = -{\rm Im} \left ( \Sigma_{n,\pm}(\tilde{E})
\right ) $ is related to the lifetime of the 2DEG electrons to scatter
in a spin-dependent way off the ferromagnet.  This new spin-dependent
scattering channel will result in different conductivities for the two
spin channels, and will be addressed in
Sec. \ref{spin_dependent_transport}.

\section{Results for Relevant Systems}
\label{coupling_results}
We have shown how to calculate the spin-dependent effects for the
coupled 2DEG-ferromagnet system.  In this section we apply this method
to two experimentally realizable systems: the silicon inversion layer
and the naturally-formed InAs surface layer.  Silicon MOSFET-type
devices are ubiquitous in present technology and any spintronics
device in this system would have a strong industrial base.  The
inversion layer is created by a gate bias, which increases the
coupling with the ferromagnet by pressing the electrons close to the
interface with the (ferromagnetic) gate.  The need for ultra-thin
oxide barriers (to increase the coupling) implies that 2DEG electrons
can irreversibly tunnel into the ferromagnet. We shall account for the
effects of the current leakage.  Since the removal of the oxide
barrier (provided that it is not replaced by a Schottky barrier)
brings simplicity, we also examine the naturally occurring InAs
accumulation layer which forms an ohmic contact with the metal.  The
2DEG forms at the interface without the need for a gate bias, so the
leakage of electrons is not a problem in this system.  However, the
confinement of electrons in such a surface layer is quite weak.

\begin{figure}[t!]
\includegraphics[width=8cm]{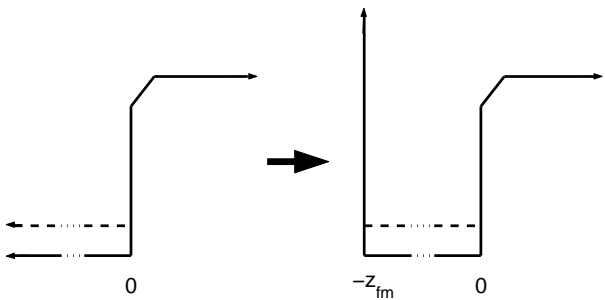}
\caption{The ferromagnet subsystem, approximated as a large but finite
box.  It consists of the exchange split ferromagnet potential, the
barrier potential, and is zero for $ z > z_{\text{b}} $.  The
wavefunction must vanish at the left boundary $ z=-z_{\text{fm}} $.}
\label{fm_subsystem}
\end{figure}

We start with calculation of the ferromagnet subsystem wavefunctions
and energy eigenvalues; we will show below how to treat the
semiconductor subsystem. The ferromagnet subsystem is made of the
ferromagnet potential for $ z<0 $, the barrier potential for $ 0 < z <
z_{\text{b}} $, and is zero for $ z > z_{\text{b}} $.  All states are
included in the calculation that decay exponentially for $ z > z_{\text{b}} $ 
( all states with energy less that zero).  To fix the
normalization of the ferromagnet states and the ferromagnet density of
states, the ferromagnet continuum is approximated by a large but
finite box (see Fig. \ref{fm_subsystem}).  Convergence is checked for
both the size of the box and the number of points in the box.

The ferromagnet has the same parameters for all calculations below.
The exchange-split parabolic bands have an effective mass equal to the
free electron mass $ m_0 $.  The majority wavevector is $ k_+^{\text{fm}}
= 1.1~{\text{\AA}}^{-1} $ and the minority wavevector is $ k_-^{\text{fm}} =
0.42~{\text{\AA}}^{-1} $,\cite{slon} corresponding to a majority Fermi
energy of $ U_{\text{fm}} = 4.6 $ eV and an exchange energy of $ \Delta =
3.9 $ eV.  The ferromagnet is assumed to be dirty, so that the
scattering is significant.  We account for this by putting by hand an
imaginary part into the ferromagnet energy eigenvalues after
calculating the particle-in-a-box eigenstates that is the same for all
wavevectors, $ \gamma^{\text{fm}}_+ = 1.1$ eV and $ \gamma^{\text{fm}}_- =
0.8 $ eV.\cite{hong} We are left with the wavefunctions and complex
energy eigenvalues for all states in the ferromagnet with the real
part of their energy less than zero.  Below we describe separately how
we calculate the semiconductor subsystem for the two systems we
consider in this paper.

\subsection{Silicon Inversion Layer}
\label{silicon_coupling}
The 2DEG in this system is strongly inverted at the interface and the
triangular approximation to the semiconductor potential is
adequate.\cite{ando} The dielectric constant in silicon is taken as $
\epsilon_{\text{Si}} = 11.7 $, the longitudinal effective mass
responsible for the confinement is $ m^\star_{\text{Si,l}} = .91~m_0
$, and the transverse effective mass relevant to the in-plane motion
of the 2DEG is $ m^\star_{\text{Si,t}} = .19~m_0 $.  The oxide barrier
between the silicon and the ferromagnet is assumed to be $
{\text{SiO}}_2 $ with barrier height (from the ferromagnet Fermi
level) $ U_{\text{SiO}_2} = 3.2 $ eV, effective mass $
m^\star_{\text{SiO}_2} = .3~m_0 $, and dielectric constant $
\epsilon_{\text{SiO}_2} = 3.9 $.\cite{si02_mass} We assume that the
electric field in the barrier, which is set by the bias between the
silicon substrate and the ferromagnetic gate, is near its breakdown
value of $ E_{\text{SiO}_2} = 12 $ MV/cm.  The electric field in the
silicon responsible for the 2DEG confinement is then $ E_{\text{Si}} =
(\epsilon_{\text{SiO}_2}/ \epsilon_{\text{Si}}) \cdot E_{\text{SiO}_2}
$.  The density in the 2DEG is assumed to be $10^{12}~{\text{cm}}^{-2}
$, corresponding to a Fermi energy of $\approx 6.3$ meV.

\begin{figure}[t!]
\includegraphics[width=8cm]{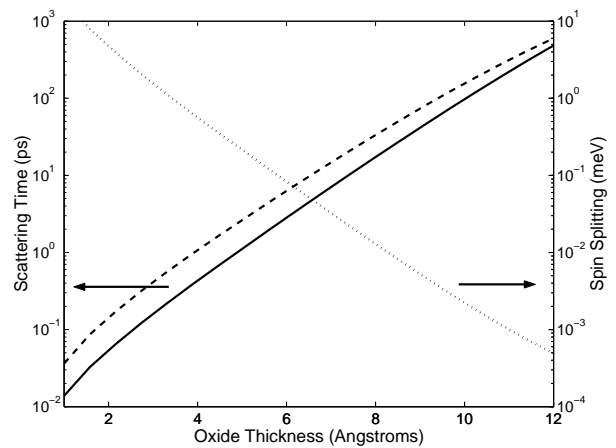}
\caption{A plot of the spin splitting $ | \Delta_+ - \Delta_- | $
(dotted line running from upper left to lower right) and the
spin-dependent scattering times $ \tau_+ $ (full line running from
lower left to upper right) and $ \tau_- $ (dashed line running from
lower left to upper right) as a function of the thickness of the
SiO$_2$ barrier for a silicon inversion layer.  The parameters used
are explained in the text.} \label{figsi_scat}
\end{figure}

The numerical results for the spin-splitting $ | \Delta_+ - \Delta_- |
$ (dotted line running from upper left to lower right) and the
scattering times for the two spin channels $ \tau_+ $ (full line
running from lower left to upper right) and $ \tau_- $ (dashed line
running from lower left to upper right) are shown in
Fig. \ref{figsi_scat} for $ E_{\text{SiO}_2} = 10 $ MV/cm.  The
horizontal axis is the thickness of the oxide barrier; sizable
coupling between the 2DEG and the ferromagnet occurs only for
ultrathin oxides at the limit of current device fabrication
techniques.  The spin-splitting is very small even for the thinnest
oxide considered; at $ 6~{\text{\AA}} $ barrier width, the
spin-splitting of $ \approx 0.1$ meV versus the 2DEG Fermi energy of $
\approx 6.3$ meV yields a static spin polarization of the 2DEG $
\approx 1 \% $.

On the other hand, the scattering time for the two spin channels is a
promising effect of the coupling that could be used in a device.  For
$ 6~{\text{\AA}}$ barrier width, the two spin channels have scattering
times of approximately 3~ps and 6~ps; the spin-dependent coupling to the
ferromagnet has opened a new spin-dependent scattering channel for the
2DEG electrons.  To utilize this effect in a device, the new
scattering times must be comparable to the intrinsic scattering time
for the silicon inversion layer (including phonons, impurities,
defects, etc., but not including the effects of the nearby
ferromagnetic layer), which at low temperatures is near 1
ps.\cite{ando} As is evident in the plot, the scattering times for the
two spin channels approach this intrinsic value for very thin oxide
thicknesses, while their relative ratio remains the same.  Because the
two times are near the intrinsic time, but are still quite different
(more than a factor of 2), we would expect that any physical quantity
that depends on the scattering time would see this spin effect.  This
will be addressed later when we discuss 2DEG transport under the gate.

\subsection{InAs Surface Layer}
\label{inas_coupling}
The 2DEG at the InAs surface is naturally formed.  As opposed to the
strongly inverted silicon inversion layer discussed above, the
confining potential in the InAs system is quite weak and a realistic
calculation of the coupling requires that the potential be calculated
self-consistently with the density distribution in the 2DEG.  We use
the coupled Schrodinger and Poisson equations in the semiconductor
region $ z > z_{\text{b}} $,
\begin{eqnarray}
\frac{-\hbar^2}{2} \frac{\partial}{\partial z} \left (
\frac{1}{m^\star} \frac{\partial \chi^{\text{sc}}_n}{\partial z}
\right ) &+& V^{\text{sc}}(z) \chi^{\text{sc}}_n (z)
=\epsilon^{\text{sc}}_n \chi^{\text{ sc}}_n (z) \label{schrodinger}\\
\frac{\partial^2 V^{\text{sc}}}{\partial z^2}& =& \frac{-4 \pi
e^2}{\epsilon_{\text{sc}}} N_0 \sum_n | \chi^{\text{ sc}}_n(z) |^2
\label{poisson}
\end{eqnarray}
where the total density in the 2DEG is kept fixed, $ N_0 = (2
m^\star_{\text{InAs}} / \pi \hbar^2 ) \sum_n (
\epsilon^{\text{sc}}_{\text{F}} - \epsilon^{\text{sc}}_n) $.  The band
bending in the barrier is negligible.  Unlike the silicon inversion
layer, more than one subband in the InAs surface layer is occupied by
electrons.  Solving these equations simultaneously results in the 2DEG
wavefunctions $ \chi^{\text{sc}}_n(z) $, energies $
\epsilon^{\text{sc}}_n $, and the semiconductor potential $
V^{\text{sc}}(z) $.  The InAs parameters used in the calculation are
an effective mass of $ m^\star_{\text{InAs}} = .023~m_0 $, dielectric
constant $ \epsilon_{\text{InAs}} = 14.6 $, and 2DEG density $ N_0 =
10^{12}~ {\text{cm}}^{-2} $.  The barrier is taken as $
{\text{Al}}_2{\text{O}}_3 $ with a barrier height from the ferromagnet
Fermi level of $ U_{\text{Al}_2\text{O}_3} = 1.2 $~eV, effective
mass\cite{al2o3_mass} $ m^\star_{\text{Al}_2\text{O}_3} = .75~m_0 $,
and dielectric constant $ \epsilon_{\text{Al}_2\text{O}_3} = 3.9 $.

\begin{figure}[t!]
\includegraphics[width=8cm]{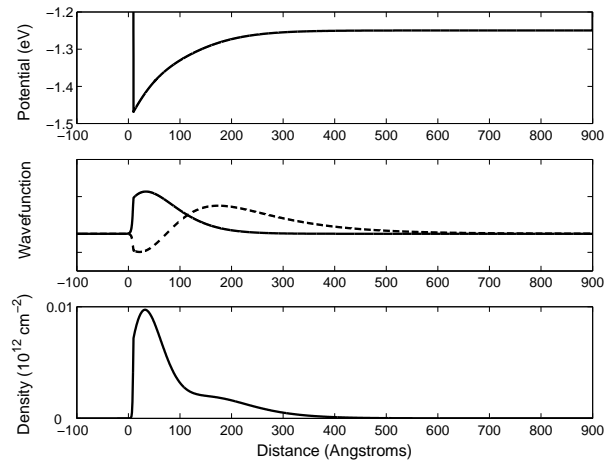}
\caption{(a) The calculated potential (eV) for the InAs surface layer
with the parameters given in the text.  The zero of energy is at the
top right side of the barrier.  (b) The wavefunctions for the two
subbands (arb. units).  (c) The density distribution in the 2DEG
(arb. units).}
\label{inas_stuff}
\end{figure}

The wavefunctions and energies associated with the calculation of the
2DEG with a $ 10~{\text{\AA}} $ barrier are shown in
Fig. \ref{inas_stuff}.  As is evident in Fig.  \ref{inas_stuff}(a),
the confining potential for the 2DEG is quite shallow and weak; the
depth of the potential is only a few hundred meV and the potential does
not flatten out until about $300~{\text{\AA}}$ from the interface. Two
subbands are occupied up to the Fermi level.  The wavefunctions of the
two subbands are shown in Fig. \ref{inas_stuff}(b), and the density
distribution in the 2DEG is shown in Fig. \ref{inas_stuff}(c).

\begin{figure}[t!]
\includegraphics[width=8cm]{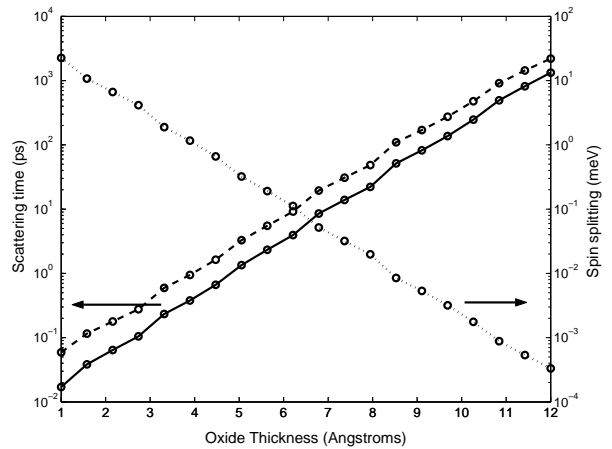}
\caption{A plot of the spin splitting $ | \Delta_+ - \Delta_- | $
(dotted line running from upper left to lower right) and the
spin-dependent scattering times $ \tau_+ $ (full line running from
lower left to upper right) and $ \tau_- $ (dashed line running from
lower left to upper right) as a function of the thickness of the
${\text{Al}_2\text{O}_3}$ barrier for the lowest subband in the InAs
surface layer.  The parameters used are explained in the text. }
\label{inas_scat}
\end{figure}

The numerical results for the coupling of the first subband of the
InAs surface layer with a ferromagnetic gate are shown in
Fig. \ref{inas_scat}, as a function of the ${\text{Al}_2\text{O}_3}$
barrier thickness.  The coupling for the second subband is an order or
magnitude weaker, and because only $ \approx 20 \%$ of the carriers
are in this subband, we neglect its presence from now on.  The spin
splitting is negligible except for intimate contact, in which case the
value of $ \approx 10$ meV means a polarization of the gas of $
\approx 15 \% $.  This value is quite high and is achievable in this
system because the barrier is unnecessary for the creation of the
2DEG.  The scattering times for the two spin channels approach the
sub-picosecond range for the thinnest barriers and intimate contact.
The intrinsic scattering time of the surface InAs 2DEG is in the
hundreds of fs range at low temperatures.\cite{watkins} The coupling
has been calculated down to an oxide thickness of $1~{\text{\AA}}$;
the tight-binding model breaks down for intimate contact.

\section{Spin-Dependent Transport Under the Ferromagnet}
\label{spin_dependent_transport}
We have calculated the spin-dependent self-energy for 2DEG states
coupled to a ferromagnetic gate, which in general results in (1) a
spin-dependent scattering time associated with the interaction of the
2DEG electrons with the ferromagnet $ \tau_\pm $ and (2) a spin
splitting ( $ | \Delta_+ - \Delta_- | $ in the single subband limit)
resulting in unequal electron densities in the two spin channels $
N_\pm $.  In this section we examine the influence of these two
results of the coupling on the in-plane transport of the 2DEG, and
find that (1) the in-plane conductivity becomes spin-dependent and (2)
2DEG electrons can irreversibly leak into the ferromagnetic gate in a
spin-dependent manner (if the ferromagnet is biased with respect to
the semiconductor).

We derive the transport equations in the 2DEG, accounting for current
leakage into the gate and, therefore, the density variation along the
semiconductor channel.  In the following we assume that the current in
the 2DEG flows in the $ \hat{x} $-direction.  The boundary condition
on the channel density will be that at the source and drain contacts
the 2DEG takes its equilibrium density related to the confinement and
the Fermi energy of the 2DEG.  The growth axis is in the $ \hat{z}
$-direction, with the ferromagnet interface at $ z = 0 $.  The
electron confinement in the $ \hat{z} $-direction is assumed to be
constant along the channel.  The system is homogeneous in the $
\hat{y} $-direction, so there is no $ y $ dependence in any of the
equations.  The two spin channels are considered to be completely
decoupled throughout the device (valid for channel lengths smaller
than the spin relaxation length).

The continuity equation for 2DEG electrons is
\begin{equation}
\frac{\partial}{\partial t} n_\pm (x,z,t) + \vec{\nabla} \cdot
\vec{j}_\pm (x,z,t) = 0~,
\label{continuity}
\end{equation}
where $ n_\pm(x,z,t) $ is the spin-dependent particle density ($
{\text{ cm}}^{-3} $) and $ \hat{j}_\pm(x,z,t) $ is the spin-dependent
particle current ($ {\text{cm}}^{-2} \cdot {\text{s}}^{-1} $).  In the
steady-state,
\begin{equation}
\frac{\partial}{\partial x} j_{x,\pm}(x,z) = - \frac{\partial}
{\partial z} j_{z,\pm}(x,z)~.
\label{steady_state}
\end{equation}
Integrating out the $ z $-dependence through the semiconductor up to
the interface at $ z = 0 $ we obtain
\begin{eqnarray}
\nonumber \frac{\partial}{\partial x} J_{x,\pm} (x) &=& -
j_{z,\pm}(x,z=0) + j_{z,\pm} (x,z=-\infty) \\ &=& - j_{z,\pm}(x,z=0)~,
\label{int_z}
\end{eqnarray}
where $J_{x,\pm}(x) = \int^0_{- \infty} dz j_{x,\pm}(x,z,)$ is the
integrated 2DEG current flowing in the $\hat{x}$-direction.  The term
$ j_{z,\pm}(x,z=0) $ represents the leakage of 2DEG electrons
irreversibly into the ferromagnetic gate and will be discussed
separately for the silicon and InAs systems below.  The term $
j_{z,\pm}(x,z=-\infty) $ represents the electrons that are injected
from the semiconductor substrate into the 2DEG.  This process is
unfavorable because electrons are injected into the 2DEG much more
efficiently from the source and drain contacts,\cite{si_leakage} so
that $ j_{z \pm}(x,z=-\infty) = 0 $.

The Drude conductivity is ordinarily
\begin{equation}
\sigma_0 = \frac{N_0 e^2 \tau_0}{m^\star}~,
\label{sigma_0}
\end{equation}
where $ N_0 $ is the 2DEG density (${\text{cm}}^{-2}$), $ -e $ is the
electron charge, $ \tau_0 $ is the intrinsic lifetime, and $ m^\star $
is the effective mass of the electrons.  The coupling of the 2DEG with
the ferromagnet introduces spin-dependence to the 2DEG density $ N_0$
and lifetime $ \tau_0 $, and, hence, to the conductivity,
\begin{equation}
\sigma_\pm = \frac{N_\pm e^2 \tilde{\tau}_\pm}{m^\star}~,
\label{sigma}
\end{equation}
where $ N_\pm(x) = \int_{- \infty}^0 dz n_\pm(x,z) $.  The
spin-dependent lifetime $ \tilde{\tau}_\pm $ includes both the
intrinsic 2DEG lifetime $ \tau_0 $ and the spin-dependent scattering
time associated with the ferromagnet $ \tau_\pm $,
\begin{equation}
\tilde{\tau}_\pm = \left ( \frac{1}{\tau_0} +
\frac{1}{\tau_\pm} \right )^{-1}~.
\label{tau_tilde}
\end{equation}
The spin dependence in the conductivity will affect the transport of
the 2DEG only if the spin-dependent densities $ N_\pm $ are very
different or the spin-dependent scattering times $ \tau_\pm $ are very
different and are comparable to the intrinsic 2DEG lifetime $ \tau_0
$.

Using this conductivity, the in-plane charge current is
\begin{equation}
-e J_{x,\pm}(x) = \sigma_\pm (x) E_x + e D_\pm(x)
\frac{\partial}{\partial x} N_\pm (x)~. 
\label{current}
\end{equation}
The diffusion constant $D_\pm(x)$ can be related to the conductivity
using the Einstein relation, and at finite temperature in two
dimensions is
\begin{equation}
D_\pm(x) = \frac{\epsilon^{\text{F}}_\pm(x) \tilde{\tau}_\pm}{m^\star}
\left ( 1 - e^{-\epsilon^{\text{F}}_\pm(x)/k_{\text{B}}T} \right )^{-1}~,
\end{equation}
where $ \epsilon^{\text{F}}_\pm(x) = \frac{\hbar^2}{2 m^\star} \left [ 4
\pi N_\pm(x) \right ] $ is the chemical potential at $ T = 0 $.  The
in-plane field, $E_x$, results form a source-drain bias and is
approximately constant throughtout the 2DEG (we do not consider the
feedback of the 2DEG density variation on the in-plane field, i.e., by
coupling them through the Poisson equation).  The in-plane charge
current Eq.~(\ref{current}) and the the current leakage
Eq.~(\ref{int_z}) are to be solved jointly to specify completely the
spin-dependent density profile $ N_\pm(x) $ along the channel.  We do
this separately for the two systems under consideration.

\subsection{Silicon Inversion Layer}
\label{si_transport}
A gate bias is necessary to create the inversion layer in silicon
MOSFET-style device.  We assume that the normal metal gate is replaced
with a ferromagnet, so that the ferromagnet is biased with respect to
the silicon substrate.  The Fermi level in the ferromagnet is lower
than the Fermi level in the semiconductor.  Hence 2DEG electrons can
tunnel into the ferromagnetic gate, inelastically fall to the Fermi
level in the ferromagnet, and have no way to get back into the
semiconductor.  This causes a leakage of the 2DEG density, so that the
current is not constant along the channel,
\begin{equation}
\frac{\partial}{\partial x} J_{x,\pm} = -j_{z,\pm}(x,z=0) =
\frac{-N_\pm(x)}{\tau_\pm}~.
\label{si_leakage}
\end{equation}
Combining Eq.~(\ref{si_leakage}) with Eq.~(\ref{current}), we have the
differential equation specifying the spin-dependent channel density:
\begin{equation}
\frac{\partial}{\partial x} \left ( D_\pm(x) \frac{\partial N_\pm}
{\partial x} \right ) + \frac{e E_x \tilde{\tau}_\pm}{m^\star
_{\text{Si,t}}} \frac{\partial N_\pm}{\partial x} -
\frac{N_\pm(x)}{\tau_\pm} = 0~.
\label{si_diff_eq}
\end{equation}
The spin-dependent scattering times $\tau_\pm$ depend on the thickness
of the oxide barrier under consideration.  We have assumed that,
except for at the source and drain, the system is homogeneous in $ x $
and that the lifetimes and scattering times are independent of the
density variation throughout the channel.  

Because the diffusion
constant is dependent upon the density in a complicated way,
Eq.~(\ref{si_diff_eq}) must be solved numerically.  In the limit where
the diffusion constant does not depend on $x$, the solution under a
single gate is
\begin{equation}
N_\pm(x) = A_\pm e^{-x/l^{\text{down}}_\pm} + B_\pm
e^{x/l^{\text{up}}_\pm},
\label{si_N}
\end{equation}
where the {\it downstream} and {\it upstream} lengths
\cite{flatte1,flatte2} are
\begin{eqnarray}
l^{\text{down}}_\pm &=& \frac{-e E_x \tau_\pm \tilde{\tau}_\pm} {2
m^\star_{\text{Si,t}}} \left ( \sqrt{1+\frac{4 D_\pm
(m^\star_{\text{Si,t}} )^2} {(eE_x)^2 \tau_\pm \tilde{\tau}_\pm^2}} +
1 \right ) \label{downstream}\\ l^{\text{up}}_\pm &=& \frac{-e E_x
\tau_\pm \tilde{\tau}_\pm} {2 m^\star_{\text{Si,t}}} \left ( \sqrt{1+
\frac{4 D_\pm (m^\star_{\text{Si,t}} )^2 } {(eE_x)^2 \tau_\pm
\tilde{\tau}_\pm^2}} - 1 \right )
\label{upstream}
\end{eqnarray}
and $ A_\pm $ and $ B_\pm $ are fixed by the boundary conditions.
Note that in all calculations in this paper, the in-plane electric
field $E_x$ was taken as negative, so that $l^{\text{up}}_\pm$ and
$l^{\text{down}}_\pm$ are defined to always be positive numbers.  The
real solution, which is calculated numerically with $D_\pm(x)$
varying along the channel, is more complicated, but the above
approximation captures the qualitative aspects of the channel density.
The density starts at its equilibrium value near the source and the
drain, which act as reservoir's for electrons.  The electrons leak
into the gate, causing the density to decrease as they move further
from the source or drain.  In the absence of a source-drain bias, the
decay of the density would be symmetric about the center of the gate;
a finite source-drain bias breaks the symmetry, and hence gives the
two decay lengths $l^{\text{up}}_\pm$ and $l^{\text{down}}_\pm$.  The
spin-dependent channel density $ N_\pm(x) $ determines the
spin-dependent channel current via Eq.~(\ref{current}).

\begin{figure}[t!]
\includegraphics[width=8cm]{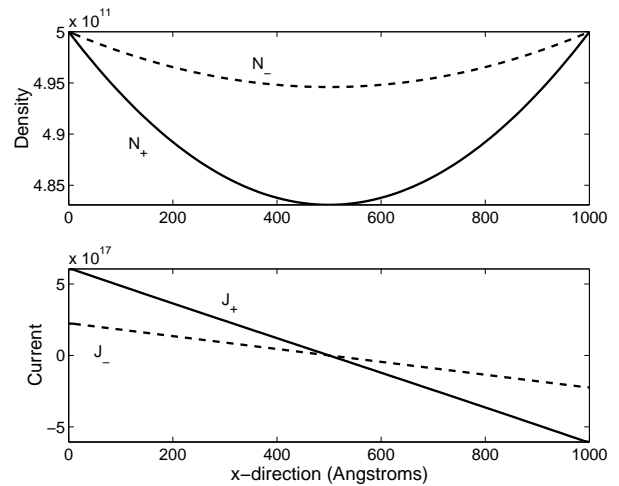}
\caption{The spin-dependent 2DEG density ($ \text{cm}^{-2} $) and
spin- dependent 2DEG current ($ \text{cm}^{-1}~\text{s}^{-1} $) under
a single ferromagnetic gate for the silicon inversion layer for {\it
zero} source-drain bias (in-plane field of $ E_x = 0 $ V/cm).  There
is a diffusion current flowing from both the source and the drain into
the gate.  The other parameters used are explained in the text.}
\label{si_backflow_fig}
\end{figure}

\begin{figure}[t!]
\includegraphics[width=8cm]{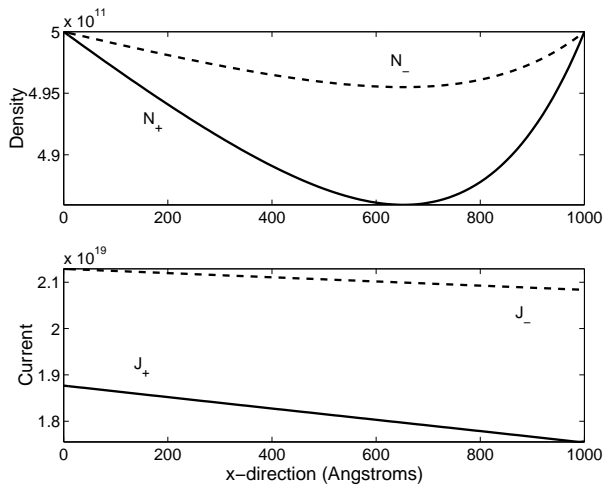}
\caption{The spin-dependent 2DEG density ($ \text{cm}^{-2} $) and
spin- dependent 2DEG current ($ \text{cm}^{-1}~\text{s}^{-1} $) under
a single ferromagnetic gate for the silicon inversion layer for an
extremely high in-plane field of $ E_x = -5000 $ V/cm.  The other
parameters used are explained in the text.}
\label{si_5000_fig}
\end{figure}

The 2DEG density and current for the silicon inversion layer coupled
to a ferromagnetic gate are shown in Fig. \ref{si_backflow_fig} and
Fig. \ref{si_5000_fig} for the low-field regime $ E_x = 0 $ V/cm and
the high-field regime $ E_x = -5000 $ V/cm respectively.  The source
is at $ x_{\text{L}} = 0 $, the drain is at $ x_{\text{R}} =
1000~{\text{\AA}} $, and both spin channels take their equilibrium
density at the source and drain, $ N_\pm(x_{\text{L}}) =
N_\pm(x_{\text{R}}) = 0.5 \cdot 10^{12}~{\text{cm}}^{-2} $.  The oxide
is assumed to be $6~{\text{\AA}}$ thick, so that the spin-dependent
scattering times are $\tau_+ =$ 4~ps and $\tau_- =$ 11~ps.  The
intrinsic scattering time is taken as 1 ps, and the temperature is 10
K.  The spin-dependent density and current were calculated by solving
the linear system of equations that results from the finite-differencing
of the differential equation specifying the density (Eq.~\ref{si_diff_eq}).  The
solution was iterated until both the density $N_\pm(x)$ and the diffusion constant $D_\pm(x)$
converged.  This method was used to calculate the current and density
for the remainder of the paper.

The effects of the leakage of channel carriers is immediately
evident in Fig. \ref{si_backflow_fig}.  Carriers under the gate can
irreversibly leak into the gate.  A diffusion current flows from the
source and from the drain into the 2DEG to replace the leaking
electrons.  The channel density is symmetric about the center of the
device, and the current flows in opposite directions on the two sides
of the gate.  The spin-dependence due to the different scattering
times for the two spin channels is evident; the $ + $ channel, which
leaks at a faster rate, falls to a lower density in the center of the
gate, and has higher currents near the source and drain due to the 
larger gradient in the density as compared to the $ - $ channel.

As the field is turned up, the drift term in Eq.~(\ref{current})
fights against the backflow at the drain contact.  At $ \approx -200 $
V/cm, the drift current overcomes the diffusive backflow and the
current through the drain becomes positive.  Curiously, the field at
which the drift current cancels the diffusion current is spin
dependent, so that it is possible to create the following situations:
(1) the drift current cancels the diffusion current for one of the
spin channels, but not the other, so that the net current that flows
through the drain is $ 100~\% $ spin polarized, and (2) the current
for one spin channel is positive while the current in the other spin
channel is exactly opposite, so that no net current flows out of the
drain, but a {\it pure} spin current flows out the drain.  Further
study of the pure spin current in the single ferromagnetic gate
silicon system will be given in a future publication.

In Fig.~\ref{si_5000_fig}, a very high in-plane field is strong enough
to overcome the backflow from the drain; the asymmetry induced by the
strong source-drain bias is evident in the plot of the spin-dependent
density.  The transport is dominated by the drift term in Eq.~(\ref{current})
due to the high in-plane field;  the diffusion current is only a small
correction. 
More current is carried in the $-$ spin channel than the
$+$ spin channel because the lifetime for the $-$ spins, 
$ \tilde{\tau}_- $, is longer
than the lifetime for the $+$ spins, $ \tilde{\tau}_+ $.

\subsection{InAs Surface Layer}
\label{inas_transport}
The 2DEG at the surface of InAs is natural and hence no gate bias on
the ferromagnet is necessary.  This keeps the Fermi levels in the
semiconductor and in the ferromagnet equal; no electrons will leak
into the gate.  The in-plane current in each spin channel must be
conserved,
\begin{equation}
\frac{\partial}{\partial x} J_{x,\pm} = 0~.
\label{inas_no_leakage}
\end{equation}
The equation specifying the channel density is, from
Eq.~(\ref{current}) and Eq.~(\ref{inas_no_leakage}),
\begin{equation}
\frac{\partial}{\partial x} \left ( D_\pm(x) \frac{\partial N_\pm}
{\partial x} \right ) + \frac{e E_x \tilde{\tau}_\pm}
{m^\star_{\text{InAs}}} \frac{\partial N_\pm}{\partial x} = 0~.
\label{inas_diff_eq}
\end{equation}
The spin-dependent density in the limit of $D_\pm(x) = $ constant is
\begin{equation}
N_\pm(x) = A_\pm + B_\pm e^{x/d_\pm}~,
\label{inas_N}
\end{equation}
with $ A_\pm $ and $ B_\pm $ specified by the boundary conditions and
the decay length $ d_\pm = m^\star_{\text{InAs}} D_\pm/(-eE_x
\tilde{\tau}_\pm) $. The decay lengths are similar to the expressions
found in Ref.~\onlinecite{flatte1} and Ref.~\onlinecite{flatte2} for infinite
spin lifetime.  Because of the spatial dependence of the diffusion 
constant, the real solution is more complicated, but the qualitative results
still hold.  

The in-plane field drives a spin-accumulation at one of the
boundaries.  A boundary condition which specifies that the density is
the same at the source and drain would prevent this from happening,
resulting in a constant density throughout the channel.  The diffusion
term in Eq.~(\ref{current}) vanishes, so that the current is only
given by the drift term. The current carried by each spin channel must
be constant due to the absence of any 2DEG leakage.  For zero
source-drain bias, the current is thus zero in both spin channels.
This is in marked contrast with the results for the silicon case,
Fig.~\ref{si_backflow_fig}, in which a diffusion current flowed
towards the center of the gate from both the source and the drain at
zero source-drain bias. This simple behavior of the InAs surface layer
with a single ferromagnetic gate is radically altered when two gates
replace the single gate, as explained below.

\section{Spin Valve With Two Adjacent Ferromagnetic Gates}
\label{spin_valve_two_gates}
\begin{figure}[t!]
\includegraphics[width=8cm]{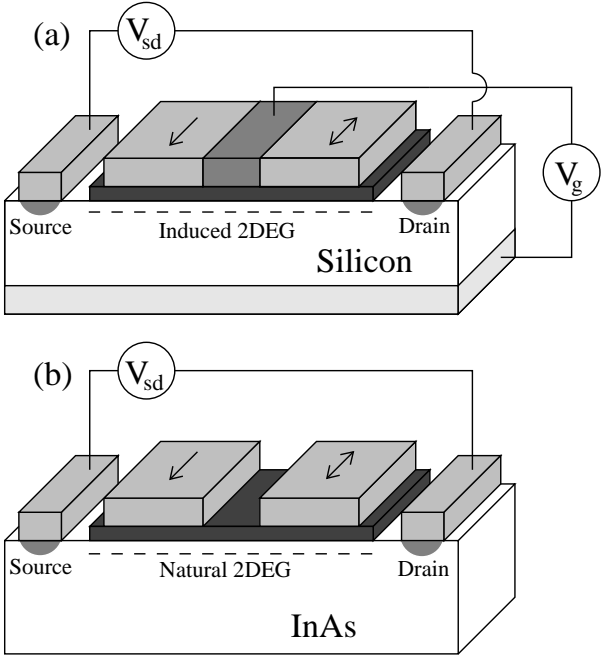}
\caption{Schematic diagrams of the spin valve proposal in (a) the
silicon and (b) the InAs systems.  The device consists of normal
nonmagnetic source and drain contacts and two ferromagnetic gates.  A
source drain bias $V_{\text{sd}}$ creates the in-plane field $E_x$.
In (a), a gate voltage $V_{\text{g}}$ is necessary to induce the
inversion in the silicon system.  The space between the ferromagnets
is filled with a nonmagnetic metal.  In (b), no gate voltage is
necessary in the InAs system. An ultrathin oxide separates the
ferromagnetic gates from the 2DEG in both systems, although the oxide
can be omitted in (b).}
\label{spin_valve_fig}
\end{figure}

Because of the general spin dependence in the 2DEG density and current
as shown above, we now discuss a simple spin-valve type device to test
the spin effects predicted by the theory.  The spin effects could not
be seen in transport experiments with a single gate and nonmagnetic
source and drain contacts.  This is because the two spin channels are
measured in parallel; switching the gate magnetization would exchange
the roles of the two spin channels but would have no effect on the
total current measured.  The single-gate spin effects could in
principle be measured through other spin-dependent means, but we
concentrate here on transport effects that would be more useful in
device considerations.

To see the spin effects in transport experiments, we proposed a simple
spin-valve design \cite{ciuti2} shown in Fig.~\ref{spin_valve_fig}.
The single gate in a normal MOSFET-style device is replaced by two
ferromagnetic gates that are very near each other.  If a gate bias is
applied (such as in the silicon system), the space between the two
ferromagnets can be filled with a nonmagnetic metal such as aluminum
in order to ensure that the 2DEG confinement is uniform between the
source and the drain.  The nonmagnetic metal between the two
ferromagnets is not necessary in systems with no gate bias (such
as the InAs system).  There is some anisotropy in the design of the
two ferromagnets (either through geometry or material) so that the
two ferromagnets have different coercive fields, which allows the switching
of the magnetization of the second ferromagnet while leaving the
magnetization of the first unchanged.  Because the two spin
channels are effectively decoupled throughout the device (due to the
long spin flip time), the addition of the second ferromagnet has a profound
influence on the total current that flows in the 2DEG.  This is
because both the density and the current in each spin channel must be
continuous throughout the device.  This matching depends on the
relative orientation of the magnetizations.  The net effect is
that the total current measured through the device will depend on the
relative orientation of the magnetization of the two ferromagnets,
which is a magnetoresistance effect. The first ferromagnet will be referred
to as the first ``gate'', and the second ferromagnet will be referred
to as the second ``gate''.

Although we propose such a device as a way to test the predictions of
the theory, the device as is could do perform the role of a memory
element.  The magnetization of the first gate remains fixed, while the
orientation of the second gate's magnetization depends on whether the
information you are trying to read out is a ``0'' or a ``1''.  The
measurement of the second gate's magnetization is read out by
measuring the current at the drain contact.  The bit could be written
using a local magnetic field (created by nearby wires, as in current
MRAM technology) that is strong enough to switch the second gate
without switching the first gate.  The two main benefits of such a
design are that (1) the nonvolatile information storage has been
incorporated onto the semiconductor, where the information processing
occurs, and (2) we take advantage of the necessity for ultrathin oxide
barriers due to the aggressive scaling of MOSFET technology to the
nanometer scale (whereas currently the ability of 2DEG carriers to
interact with the gate through an ultrathin oxide barrier is seen as a
major obstacle to be avoided).  Once demonstrated, other device
designs can be explored which do not rely on analogy to existing
structures to fully benefit from the new spin effects.

Below we discuss the spin-valve for the silicon and InAs systems
separately.  To model the 2DEG density and current throughout the
spin-valve, we assume the following.  The source is at $ x_{\text{L}}
= 0 $ and the drain is at $ x_{\text{R}} = 2200~{\text{\AA}} $.  Both
gates are $ 1000~{\text{\AA}} $ wide and the gap between the two gates
is $ 200~{\text{\AA}} $, so that the left side of the gap is at $
x_{\text{A}} = 1000~{\text{\AA}} $ and the right side of the gap is at
$ x_{\text{B}} = 1200~{\text{\AA}} $. 

The spin-dependent density with parallel gate magnetization is $
N_\pm^{\text{p}}(x) $, where $ \pm $ refers to the electron's spin
under with respect to the first gate and p refers to parallel.  The
density and current must be continuous throughout the device. 
The spin-dependent scattering times are no longer constant along
the channel due to the addition of the second gate, so that for
parallel gate magnetizations,
\begin{equation}
\tau^{\text{p}}_\pm(x) =  
\left \{
\begin{array}{ll}
\tau_\pm~, & x_{\text{L}} < x < x_{\text{A}} \\
\infty~,   & x_{\text{A}} < x < x_{\text{B}} \\
\tau_\pm~, & x_{\text{B}} < x < x_{\text{R}} \\
\end{array}
\right  .~,
\label{tau_p}
\end{equation}
and the spin-dependent lifetimes are
\begin{equation}
\tilde{\tau}^{\text{p}}_\pm(x) =  
\left \{
\begin{array}{ll}
\tilde{\tau}_\pm~, & x_{\text{L}} < x < x_{\text{A}} \\
\tau_0~,   & x_{\text{A}} < x < x_{\text{B}} \\
\tilde{\tau}_\pm~, & x_{\text{B}} < x < x_{\text{R}} \\
\end{array}
\right .~.
\label{tilde_tau_p}
\end{equation}
$\tilde{\tau}_\pm$ is specified by Eq.~(\ref{tau_tilde}).
The density and current in both the silicon and InAs systems can now
be calculated for parallel magnetization.

For antiparallel gate magnetizations, the second gate
magnetization is flipped with respect to that of the first gate, so the
spin-dependent scattering times for antiparallel gate configuration
are
\begin{equation}
\tau^{\text{ap}}_\pm(x) =  
\left \{
\begin{array}{ll}
\tau_\pm~, & x_{\text{L}} < x < x_{\text{A}} \\
\infty~,   & x_{\text{A}} < x < x_{\text{B}} \\
\tau_\mp~, & x_{\text{B}} < x < x_{\text{R}} \\
\end{array}
\right  .~,
\label{tau_ap}
\end{equation}
and the spin-dependent lifetimes are
\begin{equation}
\tilde{\tau}^{\text{ap}}_\pm(x) =  
\left \{
\begin{array}{ll}
\tilde{\tau}_\pm~, & x_{\text{L}} < x < x_{\text{A}} \\
\tau_0~,   & x_{\text{A}} < x < x_{\text{B}} \\
\tilde{\tau}_\mp~, & x_{\text{B}} < x < x_{\text{R}} \\
\end{array}
\right .~.
\label{tilde_tau_ap}
\end{equation}
The obvious difference from the parallel case is the exchange of the
roles of spin $+$ and $-$ under the second gate ( $\tau_\pm
\rightarrow \tau_\mp$ and $\tilde{\tau}_\pm \rightarrow
\tilde{\tau}_\mp$).  

For both parallel and antiparallel gate magnetizations the density for
both spin channels must take on its equilibrium value at the source
and drain, which we assume for both the silicon and InAs systems is
$.5 \cdot 10^{12}~{\text{ cm}}^{-2}$ for each spin channel, so that
the boundary conditions are
\begin{equation}
\left . 
\begin{array}{c}
N^{\text{p}}_\pm(x_{\text{L}}) \\ 
N^{\text{p}}_\pm(x_{\text{R}}) \\ 
N^{\text{ap}}_\pm(x_{\text{L}}) \\
N^{\text{ap}}_\pm(x_{\text{R}}) \\
\end{array}
\right \}
= 0.5 \cdot 10^{12}~{\text{cm}}^{-2}~.
\label{bc}
\end{equation}
These boundary conditions are sufficient to fully
calculate the spin-dependent density and current throughout the
spin-valve for parallel and antiparallel gate magnetizations.  
These are used next
to calculate the density and current in the two-gate spin valve
proposal for the silicon and InAs systems separately.

\subsection{Silicon Inversion Layer}
\label{si_spin_valve}
A schematic diagram of the silicon spin valve is shown in
Fig.~\ref{spin_valve_fig}(a).  The addition of the second gate causes
the scattering times and lifetimes to become dependent upon $x$.  For
parallel gate magnetizations, the differential equation that must be
solved is
\begin{equation}
\frac{\partial}{\partial x} \left ( D^{\text{p}}_\pm \frac{\partial
N^{\text{p}}_\pm} {\partial x} \right ) + \frac{e E_x}{m^\star
_{\text{Si,t}}} \frac{ \partial }{\partial x} \left
(\tilde{\tau}^{\text{p}}_\pm N^{\text{p}}_\pm \right ) -
\frac{N^{\text{p}}_\pm}{\tau^{\text{p}}_\pm} = 0~.
\label{si_p_diff_eq}
\end{equation}
where the spin-dependent scattering time and lifetime along the
channel are specified by Eq.~(\ref{tau_p}) and
Eq.~(\ref{tilde_tau_p}) respectively.  These equations must be solved
numerically as previously described, with the boundary conditions
Eq.~(\ref{bc}).

\begin{figure}[t!]
\includegraphics[width=8cm]{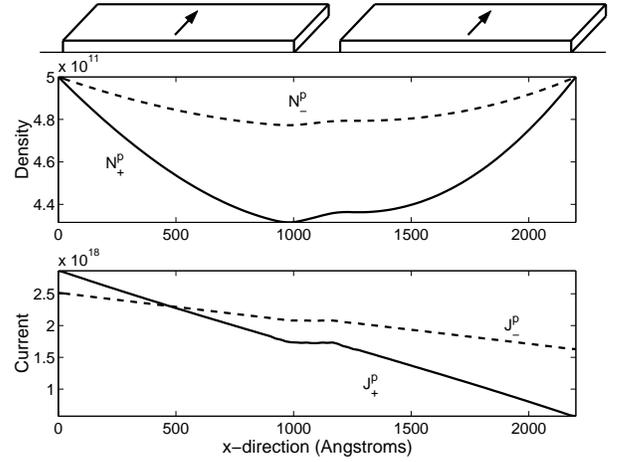}
\caption{The density ($\text{cm}^{-2}$) and current
($\text{cm}^{-1}~\text{s}^{-1}$) for a silicon inversion layer with
parallel gate magnetizations.  The parameters are explained in the
text.}
\label{si_p_500_fig}
\end{figure}

The density and current in the silicon inversion layer with parallel
gate magnetization are shown in Fig.~\ref{si_p_500_fig}.  The in-plane
field that drives the current is $ E_x = -500 $ V/cm and the
temperature is 10~K.  The intrinsic scattering time is $ \tau_0 = $
1~ps and the spin-dependent scattering times are $ \tau_+ = $ 4~ps and
$ \tau_- = $ 11~ps, consistent with a $6~{\text{\AA}}$ barrier.  The
behavior is very different compared to the single gate case,
Fig.~\ref{si_5000_fig}. The conductivity changes from under the first
gate to the gap region (likewise from the gap region to under the
second gate).  Because $ E_x $ is constant, this requires that the
density adjust itself to make both the density and current
continuous. Spin accumulation can result if the in-plane driving field
is sufficiently strong.  Note that the current in the gap region of
the device is constant because there is no electron leakage in this
region.

\begin{figure}[t!]
\includegraphics[width=8cm]{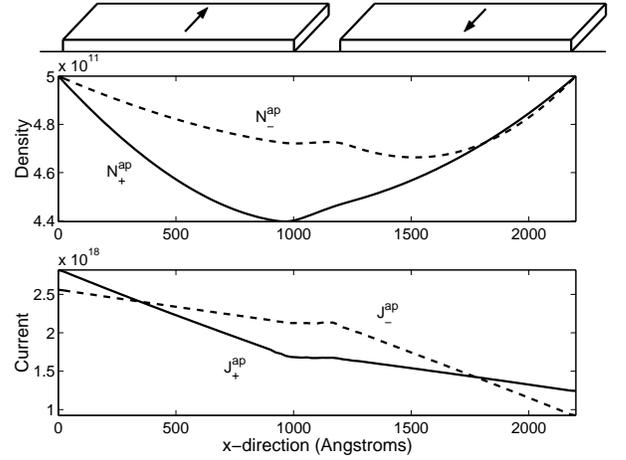}
\caption{The density ($\text{cm}^{-2}$) and current
($\text{cm}^{-1}~\text{s}^{-1}$) for a silicon inversion layer with
antiparallel gate magnetizations.  The two spin channels are labelled
with respect to the first gate.}
\label{si_ap_500_fig}
\end{figure}

For antiparallel gate magnetizations, the differential equation that
must be solved is the same as for the parallel case
(Eq.~(\ref{si_p_diff_eq})), except change p $\rightarrow$ ap in all
superscripts.  The scattering time is specified by
Eq.~(\ref{tau_ap}) and the lifetime is specified by
Eq.~(\ref{tilde_tau_ap}).  Compared to the device with parallel
magnetization, we see marked differences in both the density and
current in the antiparallel case (see Fig.~\ref{si_ap_500_fig}).
Again, this is because when crossing through the different regions of
the device, the density and current must be continuous.  Compared to
the parallel case, the $+$ spin channel now sees the conductivity and
leakage of the $-$ spin channel under the second gate, so the matching
is completely different.  The total current is different in the two
cases, causing a magnetoresistance effect.

\begin{figure}[t!]
\includegraphics[width=8cm]{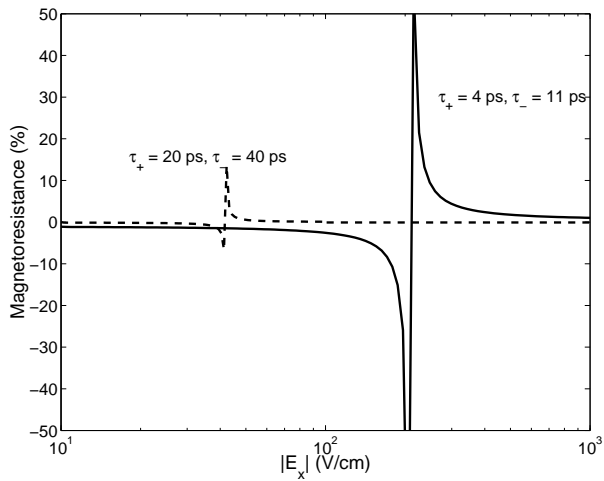}
\caption{The full line is the magnetoresistance (Eq.~(\ref{mr})) as a function of the
in-plane driving field $E_x$, using the same parameters as in 
Fig. \ref{si_p_500_fig} and
\ref{si_ap_500_fig}.  The dashed line is the magnetoresistance using scattering
times appropriate for a thicker barrier. The high-field limit of this plot
was discussed in Ref. \onlinecite{ciuti2}.}
\label{si_mr_fig}
\end{figure}

To clearly see the magnetoresistance effect caused by the switching of
the second gate's magnetization, in Fig.~\ref{si_mr_fig} we plot the 
following magnetoresistance percentage:
\begin{equation}
{\text{MR}} = 100 \cdot \frac{J_x^{\text{p}}(x_{\text{R}})-
J_x^{\text{ap}}(x_{\text{R}})} {J_x^{\text{p}}(x_{\text{R}})}~,
\label{mr}
\end{equation}
where $J_x^{\text{p}}(x_{\text{R}})$ and
$J_x^{\text{ap}}(x_{\text{R}})$ are the sum of the current in the two
spin channels with parallel magnetization and antiparallel
magnetization respectively.  The current is assumed to be measured at
the drain ($x=x_{\text{R}}$).  All parameters are the same as
previously discussed. The magnetoresistance is quite small for large
in-plane driving fields, where the drift current dominates the
transport.  The difference between the parallel and antiparallel
currents, $ J^{\text{p}}_x(x_{\text{R}})-J^{\text{ap}}_x(x_{\text{R}})
$, is small compared to $ J^{\text{p}}_x(x_{\text{R}}) $ (or $
J^{\text{ap}}_x(x_{\text{R}}) $) because of the strong drift.  As the
in-plane field is decreased, the diffusion current becomes more
important, which opposes the drift current at the drain.  The
different density profiles in the parallel and antiparallel cases
imply that the diffusion current a the drain contact is different in
the two cases; the spin effects are more pronounced and the
magnetoresistance grows.  At some critical field ($ \approx -200 $
V/cm) the drain current for parallel magnetizations equals zero; the
drift current flowing in the $ + \hat{x} $-direction is exactly
cancelled by the diffusion current flowing in the $ - \hat{x}
$-direction.  Due to the way in which Eq. (\ref{mr}) is defined, this
causes a divergence of the magnetoresistance.  As the in-plane field
is reduced further, the total current in the parallel configuration
becomes negative as the diffusive backflow overtakes the drift current
at the drain.  At low source-drain bias the drain current in both the
parallel and antiparallel configurations are negative as the backflow
completely dominates the current; a small magnetoresistance is still
present.  The spin-valve would ideally be operated just above the
critical field, where the drain current is still positive but the
effects of the diffusion current are important.  In addition, the
dashed line in the Fig.~\ref{si_mr_fig} is the same calculation using
the scattering times for a thicker oxide barrier.  The effect is still
present but moves to a lower in-plane field.  This is because there is
less leakage, and hence the diffusive backflow at the drain is lower
than for a thinner barrier.

\subsection{InAs Surface Layer}
\label{inas_spin_valve}
A schematic diagram of the InAs system is shown in
Fig.~\ref{spin_valve_fig}(b).  
For parallel magnetizations, the differenctial equation
that must be solved is
\begin{equation}
\frac{\partial}{\partial x} \left ( D^{\text{p}}_\pm
\frac{\partial N^{\text{p}}_\pm}{\partial x} \right )
+ \frac{e E_x}{m^\star_{\text{InAs}}} \frac{\partial}
{\partial x} \left ( \tilde{\tau}^{\text{p}}_\pm 
N^{\text{p}}_\pm \right ) = 0~,
\label{inas_p_diff_eq}
\end{equation}
where the scattering times and lifetimes are specified by
Eq.~(\ref{tau_p}) and Eq.~(\ref{tilde_tau_p}) and the 
boundary conditions are specified by Eq.~(\ref{bc}).

\begin{figure}[t!]
\includegraphics[width=8cm]{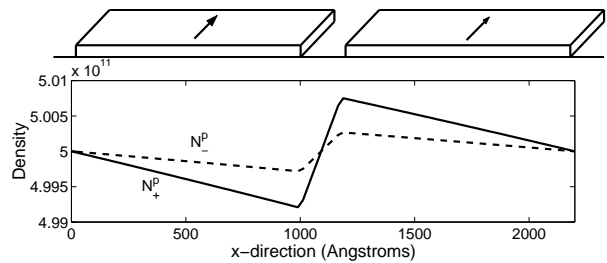}
\caption{The density (in units of ${\text{cm}}^{-2}$) 
for an InAs surface layer with parallel gate
magnetizations.  The parameters are explained in the text.}
\label{inas_p_500_fig}
\end{figure}

The density in the InAs surface layer with parallel gate
magnetization are shown in Fig.~\ref{inas_p_500_fig}.  The in-plane field that
drives the current and the boundary conditions on the source and drain
densities are the same as in the silicon case discussed previously.
The intrinsic scattering time is taken as 0.1~ps, and the
spin-dependent scattering times are taken as $\tau_+ =$ 0.3~ps $\tau_-
= $1~ps, consistent with a barrier of less than 5~\AA.  In contrast to
the single gate case, the density is not constant throughout the
device, because it must adjust itself at the interfaces between the
three regions to keep the current constant.  Both spin channels must
decrease their density in the gap region to keep the current constant,
but the density profile for the two spin channels is very different
because the lifetimes are different in the gate regions.  This process
leads to a static spin polarization at the interfaces between
different regions of the device, or spin accumulation.  This
accumulation must decay back to the equilibrium value at the drain.
The current in each spin channel is constant throughout the device 
($1.47~\cdot~10^{18}~{\text{cm}}^{-1}~{\text{s}}^{-1}$ for the 
$+$ channel, $1.74~\cdot~10^{18}~{\text{cm}}^{-1}~{\text{s}}^{-1}$ for the 
$-$ channel) due to the ``floating'' gates.

For the thinnest of oxides and intimate contact, Fig.~\ref{inas_scat}
implies that the spin splitting becomes important.  This would make
the equilibrium values for the two spin channels at the source and
drain spin-dependent.  This asymmetry further enhances the spin
effects due just to the scattering time; the qualitative picture
remains the same.

\begin{figure}[t!]
\includegraphics[width=8cm]{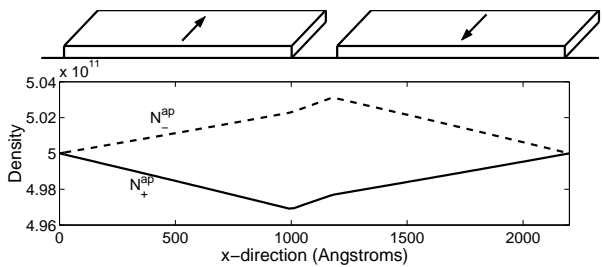}
\caption{The density $({\text{cm}}^{-2}$) 
for an InAs surface layer with antiparallel gate
magnetization.  The two spin channels are labelled with respect to the
first gate.  The parameters are explained in the text.}
\label{inas_ap_500_fig}
\end{figure}

For antiparallel gate magnetizations, the differential equation is the
same as for the parallel case (Eq.~(\ref{inas_p_diff_eq})), except 
change p $\rightarrow$ ap in all superscripts.  The scattering times
and lifetimes are given by Eq.~(\ref{tau_ap}) and Eq.~(\ref{tilde_tau_ap}).
As has already been discussed, the density and current are quite
different from the parallel case due to different matching conditions
for each of the spin channels.  The density is plotted in Fig.~\ref{inas_ap_500_fig}.  Because
of the exchanging of the roles of the two spin channels under the
second gate, the density for the $ + $ channel increases to well over
its equilibrium value in the gap region, and the $ - $ channel
decreases to well below its equilibrium value.  Again, this is another
example of spin accumulation at the interfaces between different
regions of the device.  
The current in each spin channel is constant throughout the device 
($\approx~1.6~\cdot~10^{18}~{\text{cm}}^{-1}~{\text{s}}^{-1}$ for 
both spin channels).

The magnetoresistance, Eq.~(\ref{mr}), is relatively constant at
$\approx 1~\% $ for 
all reasonable in-plane fields.  This is because the diffusion term
is always much smaller than the drift term (the spin accumulation
that occurs is always very small, much smaller than the leakage-induced
changes in the density in the silicon system), so that the magnetoresistance
is basically just specified by the difference in the spin-dependent
lifetimes. There is never any backflow, as in the silicon
case, so the divergent structure in the silicon magnetoresistance (see
Fig.~\ref{si_mr_fig}) is not seen in the InAs case.

\section{Conclusions}
\label{conc}
We presented above a comprehensive theoretical treatment of the
spin-dependent electronic and transport properties of a
two-dimensional electron gas, under the strong influence of the
proximity of a ferromagnetic layer.  By constructing an appropriate
Green`s function, we determined the complex self-energy of the quantum
confined electrons in the semiconductor, which are coupled quantum
mechanically with the spin-polarized Fermi sea in the
ferromagnet. Using a Hamiltonian coupling the two regions derived in a
tight-binding-like approach, we calculated the spin-dependent
properties of two paradigmatic systems: (i) the gate-induced inversion
layer in a ferromagnetic metal-oxide-silicon junction; (ii) the
(spontaneous) accumulation layer of InAs separated from a ferromagnet
by a thin oxide barrier.  The ferromagnetic proximity induces a
spin-splitting of the quantum confined subbands in the semiconductor
and a spin-dependent broadening, which make the in-plane transport
spin-dependent. We studied extensively the dependence of the
ferromagnetic proximity as a function of the thickness of the thin
oxide layer separating the semiconductor from the metal. Our results
show that the spin-dependent lifetime broadening is the main effect,
whereas the spin-splitting becomes sizeable only for nearly intimate
contact between the semiconductor and the ferromagnet.

Knowledge of the spin-dependent electronic properties of the
two-dimensional electron gas led to a treatment of the in-plane
transport of spin-dependent current. The leakage current into the gate
creates a density gradient along the semiconducting layer. The
resultant drift and diffusion terms of the source to drain current in
the semiconductor were treated above in a self-consistent manner with
the leakage and the density variation. While the leakage current is
usually considered a limitation for electronic field-effect
transistors with very thin oxide layers, the leakage current plays a
positive role in the spin-dependence of the transport.  We applied our
transport theory to our recently proposed spin valve with two
neighboring ferromagnetic gates. The detailed results of the
spin-dependent steady-state densities and currents for different
configurations of the magnetization in the gates yielded an explicit
understanding of dependence of the magnetoresistance as a function of
the source-drain bias. Notably, for a critical bias, a pure spin
current (i.e., with a zero net charge current) can be created at the
drain contact. This effect is caused by the gate leakage (e.g. for the
silicon inversion layer) and is due to the compensation of net drift
and diffusion currents.

We hope that the interesting new physics in the transport governed by
the proximity effect will stimulate further explorations by
experiments and by more realistic simulations, opening up the
possibility of creating field-effect spintronics devices, as an
alternative to spin injection devices.

\begin{acknowledgments}
This work is supported by DARPA/ONR N0014-99-1-1096, NSF DMR 0099572,
the Swiss National Foundation (for partial support of CC), and
University of California Campus-Laboratories Cooperation project (for
JPM).  We thank Edward T. Yu for helpful discussions.
\end{acknowledgments}

\end{document}